\documentclass[12pt,letterpaper]{article}
\usepackage{setspace}
\usepackage{color}
\usepackage{amsmath}
\usepackage{amsfonts}
\usepackage{verbatim}
\usepackage{amssymb}
\usepackage{graphicx,bm}
\usepackage{bbm}
\usepackage{cite}
\usepackage{authblk}
\usepackage{amsthm}
\usepackage{enumerate}
\usepackage{comment}
\usepackage{hyperref}
\usepackage{amssymb}
\usepackage{epstopdf}
\usepackage{epsf}
\usepackage{authblk}

\newcommand{\ima}{\mathrm{i}}

\date{}

\begin{document}

\title{Quantum Cosmology in $f(Q)$ theory}
\author[1]{N. Dimakis\thanks{nsdimakis@gmail.com; nsdimakis@scu.edu.cn}}
\author[2,3]{A. Paliathanasis\thanks{anpaliat@phys.uoa.gr}}
\author[4]{T. Christodoulakis\thanks{tchris@phys.uoa.gr}}
\affil[1]{Center for Theoretical Physics, College of Physics, Sichuan University, Chengdu 610064, China}
\affil[2]{Institute of Systems Science, Durban University of Technology, Durban 4000,
South Africa}
\affil[3]{Instituto de Ciencias F\'{\i}sicas y Matem\'{a}ticas, Universidad Austral de
Chile, Valdivia 5090000, Chile}
\affil[4]{Nuclear and Particle Physics section, Physics Department, University of
Athens, 15771 Athens, Greece}

\setcounter{Maxaffil}{0}
\renewcommand\Affilfont{\itshape\small}

\maketitle

\begin{abstract}
We use Dirac's method for the quantization of constrained systems in order to quantize a spatially flat Friedmann--Lema\^{\i}tre--Robertson--Walker spacetime in the context of $f(Q)$ cosmology. When the coincident gauge is considered, the resulting minisuperspace system possesses second class constraints. This distinguishes the quantization process from the typical Wheeler-DeWitt quantization, which is applied for cosmological models where only first class constraints are present (e.g. for models in General Relativity or in $f(R)$ gravity). We introduce the Dirac brackets, find appropriate canonical coordinates and then apply the canonical quantization procedure. We perform this method both in vacuum and in the presence of matter: a minimally coupled scalar field and a perfect fluid with a linear equation of state. We demonstrate that the matter content changes significantly the quantization procedure, with the perfect fluid even requiring to put in use the theory of fractional Quantum Mechanics in which the power of the momentum in the Hamiltonian is associated with the fractal dimension of a L\'evy flight. The results of this analysis can be applied in $f(T)$ teleparallel cosmology, since $f(Q)$ and  $f(T)$ theories have the same degrees of freedom and same dynamical constraints in cosmological studies.
\end{abstract}

\section{Introduction}

Quantum cosmology is part of a quantum theory, where the quantization process
is applied in large gravitational scales. Although quantum cosmology is expected to be
related to a reduced version of quantum gravity, the fact that we still lack a complete theory of the latter, leaves
the theoretical interpretation of quantum cosmology open to debate \cite{rev1}. The initial value problem of cosmology, the inflationary scenario, the
initial cosmological singularity, and the derivation of the quantum state of
the universe are only a few of the open problems which quantum cosmology attempts
to address \cite{sm1,sm2,sm3,sm4,sm5,sm6}.

One of the first attempts in the quantization of gravity was performed by
DeWitt \cite{deWitt} and latter by Wheeler \cite{whe}. They introduced a
canonical quantization of gravity leading to the so-called
Wheeler-DeWitt (WDW) equation. In the $3+1$ decomposition notation of
General Relativity, the WDW equation follows from the Hamiltonian constraint
\begin{equation}
\widehat{\mathcal{H}}\Psi =\left[ -4\kappa ^{2}``\mathcal{G}_{ijkl}\frac{\delta ^{2}}{%
\delta h_{ij}\delta h_{kl}}"+\frac{\sqrt{h}}{4\kappa ^{2}}\left( -\mathcal{R}[h_{ij}]%
+2\Lambda +4\kappa ^{2}T^{00}\right) \right] \Psi =0,  \label{WDW1}
\end{equation}%
where $\mathcal{G}_{ijkl}$ is the superspace metric defined as
\begin{equation}
\mathcal{G}_{ijkl}=\frac{1}{2\sqrt{h}}\left(
h_{ik}h_{jl}+h_{il}h_{jk}-h_{ij}h_{kl}\right) ,  \label{WDW2}
\end{equation}%
in the space of all three--dimensional geometries, with metric $%
h_{ij}$ and Ricci scalar $\mathcal{R}[h_{ij}]$; the quotation marks in \eqref{WDW1} are used to indicate the fact that the operator is ill defined when acting on smooth functionals. The $T^{00}$ denotes the energy density
component for the matter source. Expression (\ref{WDW1}) does not define a
differential equation in the usual sense. Heuristically, it can be thought of as a family of
differential equations at each point of a three-dimensional hypersurface. When a cosmological model is described by
a minisuperspace approximation, then - because of the additional symmetries of the
background space - the expression (\ref{WDW1}) describes a single differential
equation known also as the Schr\"{o}dinger equation of quantum cosmology. In
the case of a closed Friedmann--Lema\^{\i}tre--Robertson--Walker
(FLRW) spacetime with a quintessence scalar field as a matter source, the WDW
equation was studied independently by Hartle and Hawking in \cite{ha1} and
Vilenkin in \cite{ha2}. Since then, similar approaches have been widely applied in the
literature for many other cosmological models with a minisuperspace
description; see for instance \cite{qq1,qq2,qq3,qq4,qq5,qq6,qq7,qq8,qq9} and
references therein. Nonetheless, the canonical quantization proposal founded in the initial works by
DeWitt and Wheeler is not the unique approach towards quantum cosmology, other
attempts are based in string theory \cite{st01,st02,st02b} and in loop
quantum gravity \cite{st03,st04}. In this work we focus in the
canonical quantization approach and do not discuss further any alternative
attempts to quantum cosmology.

One major problem of modern cosmology, known as the dark energy problem, is to
determine the nature of the mechanism which describes the observable
acceleration of the late universe. A usual approach applied by cosmologists in order
to solve the dark energy problem is to modify the Einstein-Hilbert Action
Integral of General Relativity by introducing functions of geometric
invariants. In this sense a geometric description is attributed to the dark energy. The most common geometric invariants which have been introduced
by cosmologists are functions and derivatives of: the Ricci scalar \cite%
{f1,f2,f3}, the torsion scalar of teleparallelism \cite{f5}, the scalar, $Q
$, of the nonmetricity tensor \cite{f6} and many others \cite{f7,f8,f9,f10,f4,f11,f12,f13,f14,f15,f16,f17,f18,f19,f20,f21,f22,Sahoo1,Sahoo2,Sahoo3,Sahoo4}.

A specific family of modified theories are the $f-$theories; these include:
the $f\left( R\right) $ gravity \cite{f1}, where $R$ is the Ricci scalar
of the Levi-Civita connection, the $f\left( T\right) $ gravity where $T$ is
the torsion of the curvature-less Weitzenb\"{o}ck connection \cite%
{ein28,Hayashi79} and the $f\left( Q\right) $ gravity \cite{f6}. The $f\left(
R\right) $, $f\left( T\right) $ and $f\left( Q\right) $ theories are completely
distinct gravitational theories, which in general provide a different
gravitational evolution. The only exception being the case where $f$ is a linear function of its argument,
which basically recovers the dynamics of General Relativity. Another common feature of these three
theories is that, in cosmological studies, they admit a minisuperspace
description, which however have different degrees of freedom. For a nonlinear
function $f$, the $f\left( R\right) $ is
a fourth-order theory of gravity, while the $f\left( T\right) $ and $f\left(
Q\right) $ gravities are second-order theories. This is because the fundamental
invariants, $T$ and $Q$, include only first derivatives. As a result, the $f\left(
R\right) $ gravity has a scalar description, where the extra degree of
freedom - provided by the higher-order derivatives - can be attributed to a
scalar field, such that the theory is equivalent to a scalar tensor
theory. The same is not true for the $f\left( T\right) $ and $f\left(
Q\right) $ theories, for which there is not any canonical scalar field
description. In particular, for the cosmological case we study here, the scalars that are introduced lead to
second class constraints according to Dirac's theory of constrained systems \cite{Sund,Diracbook}. Systems with second class constraints have no trivial quantization. This is exactly the problem we address in this work
in order to provide a quantum description of $f\left( Q\right) $ cosmology.

The presence of second class constraints signifies the existence of redundant degrees of freedom. Based on Dirac's approach for quantizing constrained systems, we introduce the Dirac brackets in order to remove the non-physical degrees of freedom and then proceed to the canonical quantization of the system. Hence, the canonical quantization of $f(Q)$ cosmology is not based on the usual procedure of just writing a Wheeler-DeWitt equation, as it happens in cosmological configurations of General Relativity, $f(R)$ gravity and other theories with only first class constraints. A special treatment must be followed, which guarantees that the right degrees of freedom are taken into consideration. We start with the vacuum case and we generalize this approach by later taking into account various matter configurations: a minimally coupled scalar field and a perfect fluid.

The structure of this work is the following: In Section \ref{sec2} we briefly present the basic definitions of the various quantities appearing in $f\left( Q\right)$ gravity. In the case of a spatially flat FLRW background space, the field
equations can be derived by a minisuperspace Lagrangian as it is given in
Section \ref{sec3}. The field equations form a singular dynamical system,
thus, in order to pass to the Hamiltonian
formulation, the primary and secondary constraints must be determined. This
analysis is performed in Section \ref{sec4}, where we make the additional distinction of the constraints in first and second class and then define the Dirac
brackets. Section \ref{sec5}
includes the canonical quantization of the
cosmological system under study based on Dirac's approach. As simple examples, we consider two applications for the
functional form of $f\left( Q\right) $. The
cosmological scenarios with a scalar field or an ideal gas are later studied in
Sections \ref{sec6} and \ref{sec7} respectively. Finally, in Section \ref%
{con} we summarize our results and draw our conclusions.

\section{Symmetric Teleparallel Gravity}

\label{sec2}

Consider the covariant derivative $\nabla $ defined by the general
connection $\Gamma _{\phantom{\alpha}\mu \nu }^{\alpha }$
\begin{equation} \label{generalconnection}
\Gamma _{\phantom{\alpha}\mu \nu }^{\alpha }=\genfrac{\{}{\}}{0pt}{}{\alpha}{\mu\nu} +K_{\phantom{\alpha}\mu \nu }^{\alpha }+L_{\phantom{\alpha}\mu \nu }^{\alpha }
\end{equation}%
where $\genfrac{\{}{\}}{0pt}{}{\alpha}{\mu\nu} $ are the Christoffel symbols%
\begin{equation}
\genfrac{\{}{\}}{0pt}{}{\alpha}{\mu\nu} =\frac{1}{2}g^{\lambda \alpha }\left(
g_{\lambda \mu ,\nu }+g_{\nu \lambda ,\mu }-g_{\mu \nu ,\lambda }\right) ,
\end{equation}%
$K_{\phantom{\alpha}\mu \nu }^{\alpha }$ is the contorsion part
\begin{equation}
K_{\phantom{\alpha}\mu \nu }^{\alpha }=\frac{1}{2}\left( T_{\phantom{\alpha}\mu \nu }^{\alpha }+\ T_{\mu
~\nu }^{~a}+T_{\nu ~\mu }^{~a}\right) ~,
\end{equation}%
which is made with the help of the torsion $T^{\alpha}_{\phantom{\alpha}\mu\nu}$ tensor, and $L_{\phantom{\lambda}\mu \nu }^{\lambda }$ is the disformation
tensor defined by the nonmetricity $Q_{\alpha \mu \nu }=\nabla
_{\alpha }g_{\mu \nu }$, as follows
\begin{equation}
L_{\phantom{\lambda}\mu \nu }^{\lambda }=\frac{1}{2}g^{\lambda \sigma
}\left( Q_{\mu \nu \sigma }+Q_{\nu \mu \sigma }-Q_{\sigma \mu \nu }\right)
\label{disften}
\end{equation}

The fundamental geometric invariant of General Relativity is the Ricci scalar, $R$,
defined by the Levi-Civita connection $\genfrac{\{}{\}}{0pt}{}{\alpha}{\mu\nu} $. On
the other hand, in the teleparallel equivalent of General Relativity, the
fundamental invariant is the torsion $T$ defined by the contorsion $K_{\mu
\nu }^{\alpha }$. Consequently, from the nonmetricity tensor we can define
the nonmetricity scalar
\begin{equation}
Q=-g^{\mu \nu }\left( L_{\phantom{\alpha}\beta \mu }^{\alpha }L_{%
\phantom{\beta}\nu \alpha }^{\beta }-L_{\phantom{\alpha}\beta \alpha
}^{\alpha }L_{\phantom{\beta}\mu \nu }^{\beta }\right)  . \label{Qdef}
\end{equation}%
Three simple equivalent theories can be constructed with the previously mentioned objects by requiring two of them to vanish and taking the Lagrangian to be a linear expression of the third. In General Relativity, the torsion $T$ and the nonmetricity $Q$ are both zero, so one is left with the curvature as the fundamental object for the description of the spacetime. In the teleparallel theory, the torsion is the fundamental quantity with both the curvature and the nonmetricity vanishing. Finally, in the theory of symmetric teleparallelism the space-time is considered to be flat, the torsion to be zero and the general relativistic equivalent of the theory is given by the Action Integral
\begin{equation} \label{Qaction}
S_{Q}=\int \!\! \sqrt{-g}Q d^{4}x~.
\end{equation}%
The latter provides field equations equivalent to that of General Relativity and the usual Einstein-Hilbert action. The satisfaction of the flatness and tortionless conditions, $R^{\kappa}_{\phantom{\kappa}\lambda\mu\nu}=0$ and $T^{\alpha}_{\phantom{\alpha}\mu\nu}=0$ respectively, are induced by adding the curvature, $R^{\kappa}_{\phantom{\kappa}\lambda\mu\nu}$, and torsion $T^{\alpha}_{\phantom{\alpha}\mu\nu}$ tensors together with Lagrange multipliers in the integrand of \eqref{Qaction}. For simplicity, we omit these terms from expression \eqref{Qaction}; for further details regarding the full action and the variation we refer to \cite{Koivisto1,Hohmann}.

Inspired by the $f-$theories of gravity, and specifically by the $f\left(
R\right) $ and the $f\left( T\right) $ generalizations, there has been proposed the following
modified theory of gravity, known as $f\left( Q\right) $ theory, where the action is given by
\begin{equation}
S=\int \!\!\sqrt{-g}\left( -\frac{1}{2}f(Q)+\mathcal{L}_{m}\right) d^{4}x.
\label{action}
\end{equation}

The gravitational field equations follow from the variation with respect to
the metric tensor \cite{Xu}%
\begin{equation}
\frac{2}{\sqrt{-g}}\nabla _{\alpha }\left( \sqrt{-g}f^{\prime}(Q) P_{\phantom{\alpha}\mu
\nu }^{ \alpha }\right) +\frac{1}{2}f(Q)g_{\mu \nu }+f^{\prime
}(Q)\left( P_{\mu \alpha \beta }Q_{\nu }^{\phantom{\nu}\alpha \beta
}-2Q_{\alpha \beta \mu }P_{\phantom{\alpha\beta}\nu }^{\alpha \beta }\right)
=T_{\mu \nu },  \label{feqgrav}
\end{equation}%
where a prime means a total derivative with respect the $Q$, i.e. $f^{\prime
}\left( Q\right) =\frac{df\left( Q\right) }{dQ}$, while the tensor $P_{\phantom{\alpha}\mu \nu }^{\alpha }$ is given by
\begin{equation}
P_{\phantom{\alpha}\mu \nu }^{\alpha }=-\frac{1}{4}Q_{\phantom{\alpha}\mu
\nu }^{\alpha }+\frac{1}{2}Q_{(\mu \nu )}^{\phantom{(\mu\nu)}\alpha }+\frac{1%
}{4}\left( Q^{\alpha }-\tilde{Q}^{\alpha }\right) g_{\mu \nu }-\frac{1}{4}%
\delta _{\phantom{\alpha}(\mu }^{\alpha }Q_{\nu )},
\end{equation}
with the help of the two different traces
\begin{equation}
Q_{\alpha }=Q_{\alpha \phantom{\mu}\mu }^{\phantom{\alpha}\mu }, \quad \tilde{Q}%
_{\alpha }=Q_{\phantom{\mu}\alpha \mu }^{\mu }~ .
\end{equation}%

Moreover, for the matter source it follows that $T_{\mu \nu }=-\frac{2}{\sqrt{-g}}%
\frac{\delta \left( \sqrt{-g}\mathcal{L}_{m}\right) }{\delta g^{\mu \nu }}$,
while the field equations with respect to the connection are
\begin{equation}
\nabla _{\mu }\nabla _{\nu }\left( \sqrt{-g}f^{\prime}(Q)P_{\phantom{\mu\nu}\alpha
}^{ \mu \nu }\right) =0 .
\end{equation}

The diffeomorphism invariance of the theory can be used to fix the gauge so that the connection $\Gamma _{\phantom{\alpha}\mu
\nu }^{\alpha }$ is zero. This is called the coincident gauge and allows for the subsequent replacement of covariant with ordinary derivatives $\nabla _{\alpha
}\rightarrow \partial _{\alpha }$. Special care however is needed when a particular ansatz
of the metric is considered, so that the coordinate system in which it is
expressed is compatible with the gauge choice $\Gamma _{\phantom{\alpha}\mu
\nu }^{\alpha }=0$, see \cite{Zhao,Hohmann2} for more details.

The $f\left( Q\right) $ gravity is a second-order theory and it has many similarities with the $f\left( T\right) $ theory; for instance, the background field equations are the same for a cosmological configuration. So, to a large extent, our treatment here can in principle be extended to the $f(T)$ theory as well. In the
following Section, we focus our analysis in the cosmological FLRW model and its minisuperspace description.

\section{Minisuperspace Lagrangian}

\label{sec3}

In large scales the Universe is to a good extent described by the spatially flat FLRW
line element
\begin{equation}
ds^{2}=-N(t)^{2}dt^{2}+a(t)^{2}\left( dx^{2}+dy^{2}+dz^{2}\right) ,
\label{FLRW}
\end{equation}%
where $N\left( t\right) $ is the lapse function and $a\left( t\right) $ is the
scale factor. For the Hubble function, $H(t)$, in a time gauge where the lapse is given by $N(t)$, we write $H=\frac{\dot{a}}{aN}$.

As we mentioned earlier, in the coincident gauge we have $\Gamma _{\phantom{\alpha}\mu \nu }^{\alpha }=0$. Even though part of the diffeomorphism invariance has been used in order to arrive at this gauge, the parametrization invariance which is present in cosmological models in
General Relativity survives for an FLRW space-time in the context of $f(Q)$ theory. The satisfaction of the flatness condition leaves an arbitrariness, in terms of a time function, in the $L^0_{00}$ component in eq. \eqref{generalconnection}; thus, in the coincidence gauge only the $\Gamma^0_{00}$ changes due to the inhomogeneous term in its transformation under time reparametrizations. Thus, one still has the
freedom to fix the lapse function $N$ arbitrarily, for details see \cite{Koivisto1,Koivisto2,Koivisto3}.

The scalar $Q$ is written as a function of the Hubble function
\begin{equation}
Q=6\left( \frac{\dot{a}}{Na}\right) ^{2}=6H^{2},  \label{QFLRW}
\end{equation}%
and a minisuperspace Lagrangian for the gravitational part can be constructed by
introducing the definition \eqref{QFLRW} with a Lagrange multiplier as
\begin{equation}
L_{gr}=-\frac{1}{2}\sqrt{-g}f(Q)+\lambda \left( Q-6\frac{\dot{a}^2}{N^2a^2}\right)
=-\frac{N}{2}a^{3}f(Q)+\lambda \left( Q-6\frac{\dot{a}^2}{N^2a^2}\right) .
\label{gravLag}
\end{equation}%
Variation of the above Lagrangian with respect to $Q$ yields the result
for the  multiplier
\begin{equation}
\lambda =\frac{N}{2}a^{3}f^{\prime }(Q).
\end{equation}

With the use of this value for  $\lambda $ back in equation %
\eqref{gravLag}, and by additionally considering possible contribution for
matter fields $L_{m}$, we obtain the minisuperspace Lagrangian
\begin{equation}
L=-\frac{3}{N}af^{\prime }(Q)\dot{a}^{2}-\frac{N}{2}a^{3}\left(
f(Q)-Qf^{\prime }(Q)\right) +L_{m}.  \label{Lag}
\end{equation}

A similar process is followed in the case of $f(R)$ cosmology in order to
construct such a Lagrangian. The Ricci scalar curvature is
introduced as an extra field, whose correct relation to the metric
coefficients and their derivatives, is obtained with the introduction of a
Lagrange multiplier like in equation \eqref{gravLag}. However, in that
process, a kinetic term for the new field emerges in the resulting
Lagrangian. As we see in \eqref{Lag}, this is not the case here. No kinetic
term appears for $Q$ and this is the reason for the appearance of additional
constraints in the theory as we shall see in the next section.

Let us first treat the vacuum case, where $L_{m}=0$. The Euler-Lagrange
equations lead to the system
\begin{subequations}
\label{euleq}
\begin{align}
\frac{3a\dot{a}^{2}f^{\prime }(Q)}{N^{2}}-\frac{1}{2}a^{3}\left(
f(Q)-Qf^{\prime }(Q)\right) & =0,  \label{eulcon}  \\
\frac{6a\ddot{a}f^{\prime }(Q)}{N}+\frac{6a\dot{a}\dot{Q}f^{\prime \prime
}(Q)}{N}+\frac{3\dot{a}^{2}f^{\prime }(Q)}{N}-\frac{6a\dot{a}\dot{N}%
f^{\prime }(Q)}{N^{2}}-\frac{3}{2}a^{2}N\left( f(Q)-Qf^{\prime }(Q)\right) &
=0, \\
a^{2}N^{2}Q-6\dot{a}^{2}& =0 \label{eulQ}.
\end{align}

Equivalently, the above equations are written by using the Hubble function as
follows
\end{subequations}
\begin{subequations}
\label{euleqH}
\begin{align}
2Qf^{\prime }(Q)-f(Q)& =0, \label{euleqHa} \\
\left( 2\dot{H}+3 N H^{2}\right) f^{\prime }\left( Q\right) + 2 H\dot{Q}f^{\prime
\prime }(Q)-\frac{N}{2}\left( f(Q)-Qf^{\prime }(Q)\right) & =0, \\
Q-6H^{2}& =0.
\end{align}%
The last equation yields the definition of $Q$ as given by \eqref{QFLRW}.
Equations \eqref{euleq} and \eqref{euleqH} are equivalent to \eqref{feqgrav} under the ansatz %
\eqref{FLRW}. Notice that \eqref{euleqH} are trivially satisfied by $%
f(Q)\propto \sqrt{Q}$. The theory $f\left( Q\right) \propto \sqrt{Q}$ corresponds to a Lagrangian which is a total derivative, thus no gravitational theory follows. In the following analysis we shall consider $%
f\left( Q\right) \neq \sqrt{Q}$ and $f\left( Q\right) \neq f_{1}Q+f_{2}\sqrt{%
Q}$, the latter because in such a case the limit of General Relativity is recovered while here we are interested in possible modifications.

\section{Hamiltonian formulation}

\label{sec4}

In this Section we utilize the Dirac-Bergmann \cite{Dirac,AndBer} algorithm
to write the corresponding Hamiltonian of the minisuperspace Lagrangian %
\eqref{Lag}. The latter is a singular Lagrangian, i.e. the Legendre
transform is not invertible. In our case, this is caused by the lack of
velocities for the degrees of freedom $N$ and $Q$ and it implies that the
theory possesses two \emph{primary constraints}. These are the vanishing
momenta $p_{N}\approx 0$ and $p_{Q}\approx 0$ since for Lagrangian %
\eqref{Lag} we have $\frac{\partial L}{\partial \dot{N}}=\frac{\partial L}{%
\partial \dot{Q}}=0$. Notice the difference from the minisuperspace cosmology corresponding to models in General Relativity or in $f(R)$ gravity, where only one primary constraint is present, the $p_{N}\approx 0$.

The symbol \textquotedblleft $\approx $\textquotedblright\ is being used to
denote a weak equality in Dirac's theory of constrained systems \cite{Sund}.
We remind that a quantity is weakly zero if it vanishes on mass shell, but
its gradient on phase space it does not, e.g. $p_{N}=0$ on mass shell, but $%
\nabla ^{ph}p_{N}=(\partial _{N},\partial _{a},\partial _{Q},\partial
_{p_{N}},\partial _{p_{a}},\partial _{p_{Q}})p_{N}=(0,0,0,1,0,0)\neq 0$.
This property is stressed by writing $p_{N}\approx 0$ and it serves to
remember that weak equalities are not to be enforced prior to calculating Poisson
brackets. This is because, in the latter, it is the components
of $\nabla ^{ph}p_{N}$ that are of importance and not $p_{N}$ itself. The full phase space is spanned by the variables $(N,a,Q,p_N,p_a,p_Q)$ and the constraints offer a projection, after the Poisson brackets are calculated, to the physical space to where the true degrees of freedom lie. By identifying the primary constraints the process has not finished, since these may lead to secondary, the secondary to tertiary, etc.

We now put in use the theory of constrained systems to present the Hamiltonian
description of the model. For the sake of clarity we present the basic
steps, but for more details we refer the interested reader to textbooks like
\cite{Sund,Diracbook}.

The total Hamiltonian is written by adding the primary constraints with some multipliers
\end{subequations}
\begin{equation}
\begin{split}
H_{T}& =p_{a}\dot{a}-L+u_{N}p_{N}+u_{Q}p_{Q} \\
& =N\mathcal{H}+u_{N}p_{N}+u_{Q}p_{Q} .
\end{split}
\label{totHam}
\end{equation}%
In the above relation we have
\begin{equation}
\mathcal{H}=-\frac{p_{a}^{2}}{12af^{\prime }(Q)}+\frac{a^{3}}{2}\left(
f(Q)-Qf^{\prime }(Q)\right) ,  \label{Hcon}
\end{equation}%
with $p_{a}=\frac{\partial L}{\partial \dot{a}}$ being the momentum for the degree of freedom $a$ and $u_{N}$, $u_{Q}$ the
multipliers serving as the missing \textquotedblleft
velocities\textquotedblright\ of the degrees of freedom $N$ and $Q$
respectively. For consistency, the primary constraints need to be
preserved in time at least weakly. In other words, if their time derivative
is not directly zero, it must be at least equal to some combination of the
constraints. This condition is expressed as
\begin{equation}
\dot{p}_{N}\approx 0\quad \text{and}\quad \dot{p}_{Q}\approx 0.
\label{consistency}
\end{equation}%
Calculating the time derivatives of $p_{N}$ and $p_{Q}$ we obtain
respectively
\begin{subequations}
\begin{align}
\dot{p}_{N}& =\{p_{N},H_{T}\}=-\mathcal{H}, \\
\dot{p}_{Q}& =\{p_{Q},H_{T}\}=-\frac{Nf^{\prime \prime }(Q)}{12af^{\prime}(Q)^2}  \label{secgen2}
\chi ,
\end{align}%
\end{subequations}
where we have defined
\begin{equation}
\chi =p_{a}^{2}-6a^{4}Qf^{\prime}(Q)^2.  \label{chicon}
\end{equation}

Due to equation \eqref{consistency} we need to impose both $\mathcal{H}\approx 0$
and $\chi \approx 0$ as \emph{secondary constraints}. We remark that
equations \eqref{secgen2} can yield directly zero for a linear $f(Q)$
function, in which case we would not be obligated to assume $\chi \approx 0$. But, as we previously stated, here we are interested in non-linear $f(Q)$ functions.

The consistency condition of preservation in time must now be imposed on the
secondary constraints as well. Over $\mathcal{H}$ it yields
\begin{equation}
\dot{\mathcal{H}}=\{\mathcal{H},H_{T}\}=\frac{u_{Q}f^{\prime \prime }(Q)}{%
12af^{\prime}(Q)^2}\chi \approx 0,
\end{equation}%
which is weakly zero due to being proportional to $\chi \approx 0$.

The total derivative of $\chi $ gives rise to
\begin{equation}
\dot{\chi}=\{\chi ,H_{T}\}=\frac{2Np_{a}}{a}\mathcal{H}-2a^{2}\left[
3a^{2}u_{Q}f^{\prime }(Q)\left( f^{\prime }(Q)+2Qf^{\prime \prime
}(Q)\right) +2Np_{a}\left( f(Q)-2Qf^{\prime }(Q)\right) \right] ,
\label{prechi}
\end{equation}%
whose vanishing, $\dot{\chi}\approx 0$, leads to the multiplier $u_{Q}$
being fixed to the value
\begin{equation}
u_{Q}\approx -\frac{2Np_{a}\left( f(Q)-2Qf^{\prime }(Q)\right) }{%
3a^{2}f^{\prime }(Q)\left( f^{\prime }(Q)+2Qf^{\prime \prime }(Q)\right) },
\label{uQfixed}
\end{equation}%
where we have used the fact that $\mathcal{H}\approx 0$ in equation %
\eqref{prechi} and we have also assumed that $f^{\prime }(Q)+2Qf^{\prime
\prime }(Q)\neq 0,~\,$i.e. $f\left( Q\right) \neq f_{1}Q+f_{2}\sqrt{Q}$
which is the trivial case of General Relativity.

The process has thus closed without any tertiary constraints emerging from the
time consistency condition. The function $u_{Q}$ is fixed, while $u_{N}$ remains
arbitrary. We can now classify the constraints we obtained, $(p_{N},p_{Q},%
\mathcal{H},\chi )$, into \textbf{first class} and \textbf{second class}.

In the first category there belong those constraints that commute (at least
weakly) with all of the others, while in the second enter those without this
property. It is easy to see that $\{p_{N},p_{Q}\}=\{p_{N},\mathcal{H}%
\}=\{p_{N},\chi \}=0$ which makes $p_{N}$ a first class constraint. At the
same time we have for $\mathcal{H}$
\begin{align}
\{\mathcal{H},p_{Q}\}& =\frac{f^{\prime \prime }(Q)}{12af^{\prime 2}}\chi
\approx 0, \\
\{\mathcal{H},\chi \}& =\frac{6p_{a}}{a}\mathcal{H}+\frac{2p_{a}}{%
3a^{2}f^{\prime }(Q)}\chi \approx 0.
\end{align}

Because the Poisson brackets of $\mathcal{H}$ with the rest of the
constraints are at least weakly zero, due to $\mathcal{H}\approx 0$ and $%
\chi \approx 0$, the $\mathcal{H}$ is also considered as a first class
quantity.

On the other hand, we notice that
\begin{equation}
\{p_{Q},\chi \}=6a^{4}f^{\prime }(Q)\left( f^{\prime }(Q)+2Qf^{\prime \prime
}(Q)\right) \neq 0,  \label{secondclass}
\end{equation}%
which is neither zero, nor a combination of the four previously acquired
constraints. Hence, Eq. \eqref{secondclass} tells us that $p_{Q}$ and $\chi $
are second class constraints.

The quantization of systems with second class constraints is not at all trivial
\cite{Sund,Diracbook,Klauder}. The prescription of simply enforcing the
constraints as supplementary conditions on the wave function cannot be
applied in this case. Imagine for example the two second class constraints
we have, $p_Q\approx 0$ and $\chi\approx 0$. If we require that their
quantum analogues need to annihilate the wave function, i.e. $\widehat{p}_Q
\Psi =0$ and $\widehat{\chi} \Psi =0$, then inescapably $[\widehat{p}_Q,%
\widehat{\chi}]\Psi = \left(\widehat{p}_Q\widehat{\chi} -\widehat{\chi}%
\widehat{p}_Q\right)\Psi =0$, but classically we have $\{p_Q,\chi\}\neq 0$
because $p_Q$ and $\chi$ are second class constraints, thus an inconsistency
emerges. Dirac characterized degrees of freedom corresponding to second
class constraints as redundant, which must first be eliminated from the
system before proceeding to its quantization \cite{Diracbook}. A way to
achieve this is through the introduction of the Dirac brackets. The method
however is not without shortcomings. Due to ambiguities in factor ordering
other approaches have been developed as well, like the addition of extra
degrees of freedom which do not affect the dynamics but turn the second
class constraints into first ones \cite{Faddeev,Batalin}. The derivation of
the appropriate Hamiltonian however in this latter case is no simple task either.

Here, we proceed our study with Dirac's method of eliminating the second
class constraints. In what regards possible factor ordering ambiguities, as
it is stressed in \cite{Hamgauge}, the differences are proportional to the
Planck constant $\hbar$, which vanishes at the classical limit making
indistinguishable the resulting quantum theories.

We start by introducing the Dirac bracket between two phase space functions $%
A$ and $B$ \cite{Sund,Diracbook}
\begin{equation}
\{A,B\}_{D}=\{A,B\}-\{A,\xi _{M}\}\left( \Delta ^{-1}\right) _{MN}\{\xi
_{N},B\},  \label{Dbra}
\end{equation}%
where $\xi =(p_{Q},\chi )$ are the elements of the set of all second class
constraints and $\Delta _{MN}=\{\xi _{M},\xi _{N}\}$. The indexes $M,N$ here
take the values $1,2$ with $\xi _{1}=p_{Q}$ and $\xi _{2}=\chi $, the $%
\Delta ^{-1}$ in \eqref{Dbra} denotes the inverse of the matrix $\Delta $.
By construction, the Dirac bracket between two second class constraints is
zero, i.e. $\{p_{Q},\chi \}_{D}=0$. With the introduction of the Dirac
bracket to generate dynamics we can enforce the relations $p_{Q}=0$ and $%
\chi =0$ as strong equations and write the reduced Hamiltonian
\begin{equation}
H_{red}=N\mathcal{H}_{red}+u_{N}p_{N},  \label{Hamred}
\end{equation}%
where
\begin{equation}
\mathcal{H}_{red}=\frac{a^3}{2}(f(Q)-2Qf^{\prime }(Q))\approx 0  \label{conred}
\end{equation}%
and $p_{N}\approx 0$ are the remaining first class constraints. Note that %
\eqref{conred} does not imply a differential equation for $f(Q)$. The zero
on the right hand side is acquired on mass shell. That is, relation %
\eqref{conred} tells us that for every $f(Q)$ function, the solution of the
system of the Euler-Lagrange equations yields such a $Q=Q(t)$, so that the
resulting combination $f(Q)-2Qf^{\prime }(Q)$, seen as a function of $t$
after substituting $Q=Q(t)$, will always be zero. Since the relation $%
f(Q(t))-2Q(t)f^{\prime }(Q(t))=0$ is an algebraic equation in $Q(t)$ and has
no explicit dependence in $t$, it can only be satisfied if $Q(t)=$const.
(excluding always the trivial case $f(Q)\propto \sqrt{Q}$). Due to the
relation of $Q$ with the Hubble function, see \eqref{QFLRW}, we can state
that:

\textit{Every $f(Q)$ theory of a spatially flat FLRW cosmology in vacuum
results in a constant Hubble function.}

In other words, the space-time metric is either Minkowski, when $Q=0$, or a
de Sitter universe if $Q=Q_0\neq0$. The on mass shell relation \eqref{conred}
is actually the equation \eqref{euleqHa}.

Before proceeding with the quantization, let us verify that the dynamical
evolution with Dirac brackets for the Hamiltonian \eqref{Hamred} is
equivalent to that under Poisson brackets for \eqref{totHam}. It is a matter
of straightforward calculations to see that the following relations hold (in
what regards the Hamiltonian \eqref{totHam} we additionally substitute in it
the obtained fixed value of $u_Q$ given by relation \eqref{uQfixed}):
\begin{align*}
\{N,H_{red}\}_D & = \{N,H_T\}=u_N \\
\{a,H_{red}\}_D & = \{a,H_T\}+ \frac{4 N}{3 a^5 f^{\prime }(Q)
\left(f^{\prime }(Q)+2 Q f^{\prime \prime }(Q)\right)}\left(\mathcal{H}+%
\frac{\chi}{12 a f^{\prime }(Q)} \right) p_Q \\
\{Q,H_{red}\}_D & = \{Q,H_T\}+\frac{N p_a}{3 a^5 f^{\prime }(Q)
\left(f^{\prime }(Q)+2 Q f^{\prime \prime }(Q)\right)}\left(\mathcal{H}+%
\frac{\chi}{12 a f^{\prime }(Q)} \right) \\
\{p_N,H_{red}\}_D & = \{p_N,H_T\}- \frac{\chi}{12 a f^{\prime }(Q)} - \frac{%
4 p_a}{3 a^5 f^{\prime }(Q) \left(f^{\prime }(Q)+2 Q f^{\prime \prime
}(Q)\right)}\left(\mathcal{H}+\frac{\chi}{12 a f^{\prime }(Q)} \right) p_Q \\
\{p_a,H_{red}\}_D & = \{p_a,H_T\}+ \frac{N}{12a^2f^{\prime }(Q)}\chi + \frac{%
8 N p_a}{3 a^6 f^{\prime }(Q) \left(f^{\prime }(Q)+2 Q f^{\prime \prime
}(Q)\right)}\left(\mathcal{H}+\frac{\chi}{12 a f^{\prime }(Q)} \right) p_Q \\
\{p_Q,H_{red}\}_D & = \{p_Q,H_T\}+ \frac{N f^{\prime \prime }(Q)}{12 a
f^{\prime}(Q)^2} \chi + \frac{2 N p_a}{3 a^2} \mathcal{F}(Q) p_Q
\end{align*}
where $\mathcal{F}(Q)$ is a function containing $Q$, $f(Q)$ and its
derivatives up to third order. It is obvious that the difference between the
Dirac brackets and the Poisson brackets are terms which are just multiples
of quantities which are bound to be zero on mass shell, that is $\mathcal{H}$%
, $\chi$ and $p_Q$. Hence, the same dynamics are obtained. The system is
purely described now by the reduced Hamiltonian $H_{red}$ and the strong
equations $p_Q=0$ and $\chi=0$. The last two remove the redundant degrees of
freedom since in $H_{red}$, neither $p_Q$ nor $p_a$ appear (the latter has
been eliminated by using $\chi=0$). It is the Hamiltonian \eqref{Hamred}
comprised by the two remaining first class constraints $p_N\approx 0$ and $%
\mathcal{H}_{red}\approx 0$ which is to be quantized.

\section{Quantization}

\label{sec5}

For the canonical quantization we need to introduce a mapping $\{\;,\;\}_D
\rightarrow -\mathrm{i} [\;,\;]$ (we work in the units $\hbar=1$). Thus, it
is the Dirac bracket which becomes the starting point of applying the
canonical quantization procedure. Dirac brackets in general do not give the
same fundamental relations among the basic coordinates that Poisson brackets
do. For example it is easy to see that in our case definition \eqref{Dbra}
yields
\begin{equation}  \label{exampleDbra}
\{Q,a\}_D = -\frac{p_a}{3 a^4 f^{\prime }(Q) \left(f^{\prime }(Q)+2 Q
f^{\prime \prime }(Q)\right)}, \quad \{Q,p_Q\}_D=0,
\end{equation}
in contrast to what we have when using the Poisson brackets, $\{Q,a\}=0$ and
$\{Q,p_Q\}=1$ respectively. This means that in general there is no obvious
assignment of multiplicative or differential operators for the basic
variables, e.g. a choice $\widehat{Q}=Q$, $\widehat{p}_Q=-\mathrm{i}
\partial_Q$ makes no sense here since the second of \eqref{exampleDbra}
implies that at the quantum level we must have $[\widehat{Q},\widehat{p}%
_Q]\Psi=0$. This is one of the complications appearing in this method,
together with the factor ordering ambiguities when trying to construct a quantum
version of \eqref{exampleDbra}.

In our case however, things can be significantly simplified. First, let us
notice that the constraint $\chi=0$, with $\chi$ given by \eqref{chicon},
implies
\begin{equation}  \label{solsc}
p_a = - a^2 \sqrt{6Q} f^{\prime }(Q) .
\end{equation}
To be more precise, $\chi=0$, has two roots $p_a = \pm a^2 \sqrt{6Q}
f^{\prime }(Q)$. But, if we take into account the definition $p_a= \frac{%
\partial L}{\partial \dot{a}}=-\frac{6a \dot{a} f^{\prime }(Q)}{N}$, we see
that the sign we need in order to have real quantities is the minus and thus %
\eqref{solsc} is obtained. A key observation is that by virtue of %
\eqref{solsc} we obtain
\begin{equation}  \label{Dbracan}
\{\mp \left(f(Q)-2Qf^{\prime }(Q)\right), \pm \frac{a^3}{(6Q)^{1/2}}\}_D = -
\frac{p_a}{(6Q)^{1/2}a^2f^{\prime }(Q)} = 1 .
\end{equation}
Hence, relation \eqref{Dbracan} reveals
\begin{equation}  \label{canvar}
q = \mp \left( f(Q)-2Qf^{\prime }(Q) \right) \quad \text{and} \quad p_q =
\pm \frac{a^3}{(6Q)^{1/2}}
\end{equation}
as our basic canonical variables. Obviously we need to set the restriction $%
Q \geq 0$.

If we adopt the reparametrization
\begin{equation}  \label{newparq}
\frac{1}{F(q)^2} =\mp \sqrt{\frac{3Q}{2}} \left(f(Q)-2 Q f^{\prime
}(Q)\right)
\end{equation}
by introducing a new function $F(q)$\footnote{We can use the plus or
minus sign in the right hand side of \eqref{newparq} depending on the situation, so as to always have a real
function $F(q)$; for more details see below in the examples.}, the
reduced Hamiltonian constraint $\mathcal{H}_{red}\approx 0$ of \eqref{conred} is
written as
\begin{equation}  \label{conred2}
\mathcal{H}_{red} = - \frac{p_q}{F(q)^2} \approx 0,
\end{equation}
where we have chosen the set of canonical variables in \eqref{canvar} for
which $q\geq 0$ when the $F(q)$ of \eqref{newparq} is real. Note that the
expression \eqref{conred2} does not imply $p_q\approx 0$ since this
corresponds through \eqref{canvar} to a zero scale factor. The latter is a
trivial solution of the Euler-Lagrange equations \eqref{euleq}, which
however makes singular the spacetime metric and thus it is not acceptable.
The satisfaction of \eqref{conred2}, in the parametrization we introduced,
tells us that the classical solution lies in the $F(q)\rightarrow \infty$ limit,
as we are going to verify in the specific examples to follow.

After having obtained the canonical variables, we are ready to apply the
mappings
\begin{equation}
p_{N}\mapsto \widehat{p}_{N}=-\mathrm{i}\partial _{N},\quad p_{q}\mapsto
\widehat{p}_{q}=-\mathrm{i}\partial _{q}
\end{equation}%
while the $\widehat{N}$ and $\widehat{q}$ are just set to act
multiplicatively. According to Dirac's approach we need to enforce the
quantum versions of the two first class constraints on the wave function, $%
\widehat{p}_{N}\Psi =0$ and $\widehat{\mathcal{H}}_{red}\Psi =0$. The first
simply implies that the wave function does not depend on $N$ explicitly, so
we have $\Psi =\Psi (q)$. In what regards the second we need first to
address the factor ordering problem. In order to do this we write the most
general linear first order differential operator which is Hermitian under a
measure $\mu (q)$, given a vanishing condition on the border for the wave
function:
\begin{equation}
\widehat{\mathcal{H}}_{red}=-\frac{\mathrm{i}}{2\mu (q)}\left[ \frac{\mu (q)%
}{F(q)^{2}}\partial _{q}+\partial _{q}\left( \frac{\mu (q)}{F(q)^{2}}%
\;\right) \right] .  \label{Hquantvac}
\end{equation}

Thus, we have to solve the simple ordinary differential equation $\widehat{%
\mathcal{H}}_{red}\Psi =0$
\begin{equation}
\frac{\Psi (q)F^{\prime }(q)}{F(q)^{3}}-\frac{\Psi (q)\mu ^{\prime }(q)}{%
2F(q)^{2}\mu (q)}-\frac{\Psi ^{\prime }(q)}{F(q)^{2}}=0,
\end{equation}%
which is satisfied by
\begin{equation}
\Psi (q)=C\frac{F(q)}{\sqrt{\mu (q)}},  \label{psigen}
\end{equation}%
where $C$ is the constant of integration serving for normalization purposes.

The solution \eqref{psigen} is such that makes the choice of a measure for
the probability amplitude irrelevant; Consider the inner product
\begin{equation}
\langle \Psi|\Psi \rangle = \int_\alpha^\beta\!\! \mu \Psi^* \Psi dq = |C|^2
\int_\alpha^\beta\!\! F(q)^2 dq,
\end{equation}
the probability amplitude can be expressed without having to adopt a
particular $\mu(q)$. What is more, any square integrable function $F \in
L^2[\alpha,\beta]$ can be used to write a formal probability. At the
classical level of course we know that $F\rightarrow \infty$ which means
that the maximum of the probability amplitude $\mu\Psi^* \Psi=F^2$ lies on the
classical trajectory. Classically, the $q$, which is associated
to $Q$ is bound to be a particular constant number, but at the quantum level we may choose it
ranging in a region $q \in [\alpha,\beta]$ (it does not necessarily have to
be bounded, we may consider $\alpha$, $\beta$ to be infinite).

We mentioned that operator \eqref{Hquantvac} is generally Hermitian given a
vanishing wave function on the boundary. However, by virtue of the solution %
\eqref{psigen} of our case, it is easy to see that the Hermiticity condition
is satisfied without any reference to a particular boundary condition. A
simple calculation yields
\begin{equation}
\langle \Psi |\widehat{\mathcal{H}}_{red}\Psi \rangle =\int \!\!\mu \Psi
^{\ast }\left( \widehat{\mathcal{H}}_{red}\Psi \right) dq=\langle \widehat{%
\mathcal{H}}_{red}\Psi |\Psi \rangle -\mathrm{i}\left[ \frac{\mu }{F(q)^{2}}%
\Psi ^{\ast }\Psi \right] _{\alpha }^{\beta }.
\end{equation}%
Substitution of \eqref{psigen} into the boundary term of the previous
relation results in $\left[ \frac{\mu }{F(q)^{2}}\Psi ^{\ast }\Psi \right]
_{\alpha }^{\beta }=\left[ |C|^{2}\right] _{\alpha }^{\beta }=0$ and hence $%
\langle \Psi |\widehat{\mathcal{H}}_{red}\Psi \rangle =\langle \widehat{%
\mathcal{H}}_{red}\Psi |\Psi \rangle $.

We proceed by considering a few pedagogical examples to see how the theory
works.

\subsection{Model 1: $f(Q)=Q^{\protect\lambda }$}

As a first example we use the function $%
f(Q)=Q^\lambda$, which serves the purpose of a simple demonstration of the
theory. The classical solution corresponds to $a(t)=$const.$\Rightarrow Q=0$%
, as long as $\lambda \geq 1$. For $\lambda<1$ there is no classical
solution except from the trivial case $\lambda=\frac{1}{2}$. For the quantum
study however we need not restrict $\lambda$ according to the classical
solution and initially we may consider that it can be any real number.

The new variable $q$ is given, according to \eqref{canvar}, by
\begin{equation}
q=|1-2\lambda |Q^{\lambda },  \label{qtoQex1}
\end{equation}%
where for the sector $\lambda >\frac{1}{2}$ we take the upper signs of %
\eqref{canvar}, while for $\lambda <\frac{1}{2}$ we consider the lower ones.
In this way $q$ is never negative, as also happens with $Q$.
Equation \eqref{newparq}, yields
\begin{equation}
F(q)^{-2}=\sqrt{\frac{3}{2}}|1-2\lambda |^{-\frac{1}{2\lambda }}q^{\frac{1}{%
2\lambda }+1}
\end{equation}%
This $F(q)$ function is square integrable in a finite region $q\in \lbrack
0,\alpha ]$ if $\lambda <0$, since then
\begin{equation}
\int_{0}^{\alpha }F(q)^{2}dq=-2\lambda \sqrt{\frac{2}{3}}\alpha ^{-\frac{1}{%
2\lambda }}(1-2\lambda )^{\frac{1}{2\lambda }}.
\end{equation}

For the other values of $\lambda $ or a non-finite region of integration, the
integral, and hence $\langle \Psi |\Psi \rangle $, diverges. Of course, we
need to mention that this diverging inner product is a situation which is
often encountered in the canonical quantization of cosmological
configurations. The probabilistic interpretation of the square of $\Psi $ is
also a matter under discussion. In addition to this, here we have an additional
important difference from the typical cosmological models: the wave
function is not dependent on the scale factor $a$, as happens in the usual
Wheeler-DeWitt quantization of General Relativity models, but it rather depends on $%
Q$ (through $q$ and equation \eqref{qtoQex1}) which is essentially related to the Hubble function.

For the diverging cases, we can perform a type of \textquotedblleft normalization\textquotedblright\
through defining probability ratios, since the explicit calculation of the
integral is easy in this case. We first demand that
\begin{equation}
\langle \Psi |\Psi \rangle =|C|^{2}\int_{\varepsilon }^{\alpha }F(q)^{2}dq=2%
\sqrt{\frac{2}{3}}|C|^{2}\frac{\lambda |1-2\lambda |^{\frac{1}{2\lambda }}}{%
\alpha ^{\frac{1}{2\lambda }}\varepsilon ^{\frac{1}{2\lambda }}}\left(
\alpha ^{\frac{1}{2\lambda }}-\varepsilon ^{\frac{1}{2\lambda }}\right) =1 .
\end{equation}%
Then we assume that $\varepsilon <<1$, while $\alpha $ is some maximum value that $q$
may take (if $\lambda >0$ we may consider $\alpha \rightarrow +\infty $).
From the above expression we obtain the value of the normalization constant $%
C$ which we can use in the relation for the probability density
\begin{equation}
\rho _{\Psi }(q)=\mu (q)\Psi ^{\ast }\Psi =|C|^{2}\frac{q^{-\frac{1}{%
2\lambda }-1}}{\sqrt{\frac{3}{2}}|2\lambda -1|^{-\frac{1}{2\lambda }}}=\frac{%
q^{-\frac{1}{2\lambda }-1}}{2\lambda \left( \epsilon ^{-\frac{1}{2\lambda }%
}-\alpha ^{-\frac{1}{2\lambda }}\right) }.
\end{equation}%
The probability of finding the system in a region $q\in (\epsilon ,\beta )$
where $\beta <\alpha $ is
\begin{equation}
P(\varepsilon ,\beta )=\int_{\epsilon }^{\beta }\rho _{\Psi }(q)dq=\frac{%
\varepsilon ^{-\frac{1}{2\lambda }}-\beta ^{-\frac{1}{2\lambda }}}{%
\varepsilon ^{-\frac{1}{2\lambda }}-\alpha ^{-\frac{1}{2\lambda }}}.
\label{probex1a}
\end{equation}

\begin{figure}[tbp]
\includegraphics[width=1\textwidth]{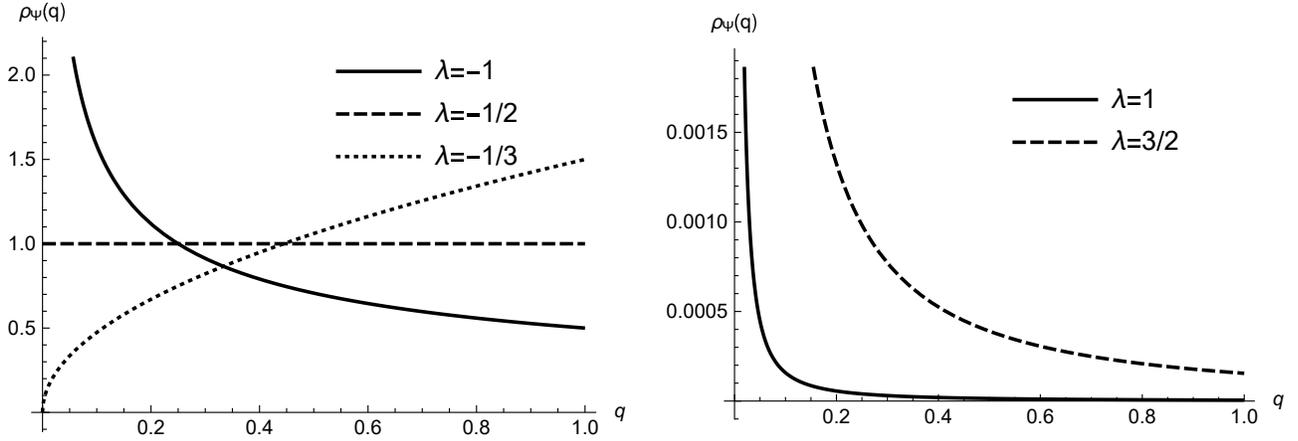}
\caption{The probability density $\protect\rho _{\Psi }(q)$ for various
values of $\protect\lambda $. In the first set of graphs, in which $\protect%
\lambda $ is negative, we have set $\protect\varepsilon =0$. The second set,
for positive $\protect\lambda $, is given for the value $\protect\varepsilon %
=10^{-10}$. In all cases we have considered, as the upper limit for $q$ is $%
\protect\alpha =1$.}
\label{fig1}
\end{figure}

For $\lambda <0$, we can extend the relation to $\varepsilon =0$ and obtain
\begin{equation}
P(0,\beta )=\left( \frac{\beta }{\alpha }\right) ^{-\frac{1}{2\lambda }}.
\end{equation}%
Since $Q$, and hence $q$, are related to the Hubble function, we can interpret
the resulting probability as one describing probable Hubble rates through $q$. See for example the Figure \ref{fig1}. For
positive values of $\lambda $ we see that the probability density sharply
peaks at $q=0$, which corresponds to $Q=0$. The latter agrees with the
classical solution of this case; remember that the classical scale factor is just a constant
and the Hubble function is zero. We also notice that larger positive values
of $\lambda $ lead to a wider possible deviation from the classical value $%
q=0$. For the negative $\lambda $, which has no classical equivalent, we see
three different behaviours: for $\lambda <-\frac{1}{2}$ the probability
peaks again at $q=0$, for $\lambda =-\frac{1}{2}$ all values of $q$ are
equiprobable, while lastly for $\lambda >-\frac{1}{2}$ the most probable
value is the one we have assumed as the upper limit for $q$.

\subsection{Model 2: $f(Q)=Q-\protect\sigma Q^{2}$}

As a second example we consider the function $f(Q)=Q-\sigma Q^{2}\,$.
In this case, for small values of $\sigma $, the general relativistic
equations are approached in the generic theory. If we assume $\sigma >0$,
the classical solution has two branches, one with $a(t)=$const. and $Q=0$
and another in which $Q=\frac{1}{3\sigma }$, which in the time gauge $N=1$
corresponds to a scale factor $a(t)=e^{\pm \frac{t}{3\sqrt{2\sigma }}}$.

The canonical variable $q$ in this case is
\begin{equation}
q=Q|3\sigma Q-1|,  \label{qtoQex2}
\end{equation}%
where again we choose the signs of \eqref{canvar} so that we have $q\geq 0$.
We notice that the value $q=0$ corresponds to either $Q=0$ or $Q=\frac{1}{%
3\sigma }$, which are the values of the classical solution. Given that, due
to the transformation \eqref{canvar} to be real, apart from $q$ we also need
$Q\geq 0$, we obtain the following possibilities by inverting \eqref{qtoQex2}%
:
\begin{equation}
Q=%
\begin{cases}
\frac{1+\sqrt{1+12\sigma q}}{6\sigma }, & \sigma >0 \\
\frac{1+\sqrt{1-12\sigma q}}{6\sigma }, & \sigma >0\text{ and }0\leq q\leq
\frac{1}{12\sigma } \\
\frac{1-\sqrt{1-12\sigma q}}{6\sigma }, & \sigma >0\text{ and }0\leq q\leq
\frac{1}{12\sigma }\text{ or }\sigma <0.%
\end{cases}%
\end{equation}%
The behaviour of the probability density is similar in the above listed
cases, thus for simplicity we analyze only the first one, in which we take $%
Q=\frac{1+\sqrt{1+12\sigma q}}{6\sigma }$. The resulting $F(q)$ function is
\begin{equation}
F(q)^{2}=\frac{2\sqrt{\sigma }}{q\left( \sqrt{12q\sigma +1}+1\right) ^{\frac{%
1}{2}}}.
\end{equation}%
Again we have a divergence at $q=0$ of the classical solution. We proceed by
normalizing in terms of ratios of probability.

First we calculate $C$ from

\begin{equation}
\langle \Psi |\Psi \rangle =|C|^{2}\int_{\varepsilon }^{+\infty }F(q)^{2}dq=1
\end{equation}%
where again we consider $\varepsilon <<1$. A straightforward computation
yields
\begin{equation}
|C|^{2}=\left[ \sqrt{\sigma }\left( \frac{4}{\left( \sqrt{12\sigma \epsilon
+1}+1\right) ^{\frac{1}{2}}}+\sqrt{2}\left( \mathrm{i}\pi +2\tanh
^{-1}\left( \frac{\left( \sqrt{12\sigma \epsilon +1}+1\right) ^{\frac{1}{2}}%
}{\sqrt{2}}\right) \right) \right) \right] ^{-1}
\end{equation}%
and the probability density is
\begin{equation}
\rho _{\Psi }(q)=|C|^{2}F(q)^{2}=|C|^{2}\frac{2\sqrt{\sigma }}{q\left( \sqrt{%
12q\sigma +1}+1\right) ^{\frac{1}{2}}}.
\end{equation}

\begin{figure}[tbp]
\centering
\includegraphics[scale=0.9]{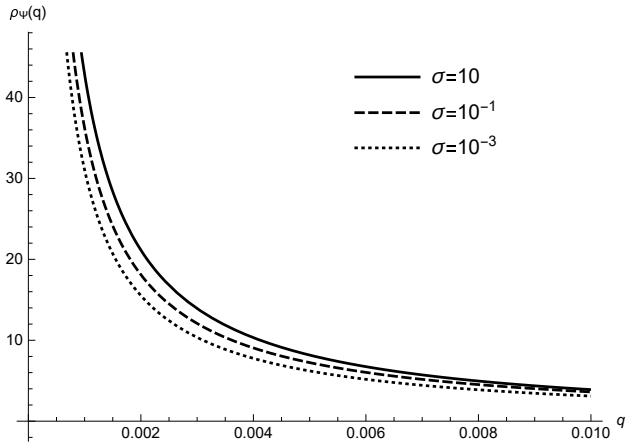}
\caption{The probability density $\protect\rho _{\Psi }(q)$ for various
values of $\protect\sigma $. As $\protect\sigma $ becomes smaller the
probability peaks more sharply near the classical solution $q=0$.}
\label{fig2}
\end{figure}

In figure \ref{fig2} we see the graph of the above function with respect to $%
q$. We observe that, as $\sigma $ becomes larger, the values of $q$ that
depart from the classical solution $q=0$ become more probable; although
there is still a sharp peak at $q=0$.

At this point we want to remark that the Minkowski and the de Sitter
classical solutions are not distinguished at the quantum level since both account for $q=0$. However, what is evident is that the classical configuration dominates the probability amplitude.

\section{The inclusion of a scalar field}

\label{sec6}

Our study up to now only referred to the vacuum case, which classically
implies a constant Hubble expansion rate. In this, and in the next, section we
expand our analysis by including matter content in order to study how it affects the
quantization procedure.

We start by considering a minimally coupled scalar field $\phi$. The matter
Lagrangian is
\begin{equation}
L_m = \frac{a^3}{2 N} \dot{\phi}^2 - N a^3 V(\phi),
\end{equation}
which enters the total Lagrangian for the system $L$ given by \eqref{Lag}.

The resulting Hamiltonian is again of the form
\begin{equation}  \label{totHamsc}
H_T = N \mathcal{H} + u_N p_N + u_Q p_Q,
\end{equation}
where this time the Hamiltonian constraint reads
\begin{equation}
\mathcal{H} = - \frac{p_a^2}{12 a f^{\prime }(Q)} + \frac{p_\phi^2}{2a^3} +
\frac{a^3}{2}\left(f(Q)-Q f^{\prime }(Q) \right) + a^3 V(\phi),
\end{equation}
with $p_\phi = \frac{\partial L}{\partial \dot{\phi}}$. The constraint
analysis is to a large extent similar as before. We have four constraints,
the two primary $p_N\approx 0$, $p_Q \approx 0$ and two secondary, $\mathcal{%
H}\approx 0$ and $\chi \approx 0$ with the latter being again the one given
by \eqref{chicon}. The conservation in time of $\chi$ results to the fixing
of the ``velocity'' $u_Q$ as
\begin{equation}
u_Q \approx \frac{N p_a \left(p_\phi^2-2 a^6 \left(f(Q)-2 Q f^{\prime }(Q)+2
V(\phi)\right)\right)}{3 a^8 f^{\prime }(Q) \left(f^{\prime }(Q)+2 Q
f^{\prime \prime }(Q)\right)} .
\end{equation}

A slight difference here is that, due to the extra presence of the scalar
field, we need to take a different linear combination of constraints as our
basis in order to reveal the maximum number of first class constraints. The
reason is that now $\{\mathcal{H},\chi\}$ is not weakly zero. However this
does not necessarily alter the number of first class constraints since it may
be an artefact of a bad choice of linear combinations of constraints; and
this is what happens in our case. If we consider as our basis of constraints
the quadruplet $\zeta_i$, $i=1,...,4$ with $\zeta_1=p_N$, $\zeta_2=p_Q$, $%
\zeta_3=\chi$, but with
\begin{equation}
\zeta_4 = \mathcal{H} - \frac{p_a \left(f(Q)-2 Q f^{\prime }(Q)+2 V(\phi
)\right)}{a^2 f^{\prime }(Q) \left(f^{\prime }(Q)+2 Q f^{\prime \prime
}(Q)\right)} p_Q \approx 0
\end{equation}
instead of just $\mathcal{H}$, then we see that $\{\zeta_3,\zeta_4\}\approx
0 $ and hence the number of first and second class constraints is the same
as before. The second class constraints are again $p_Q$ and $\chi$ (or in
our new basis $\zeta_2$ and $\zeta_3$). We choose once more the same subset
consisting of the second class constraints $\xi=(p_Q,\chi)$ and put in use
the definition of the Dirac brackets as given by \eqref{Dbra}. By invoking
the Dirac brackets for the evolution and by considering the second class
constraints as strong equations to zero we obtain the reduced Hamiltonian of
the form \eqref{Hamred}, where now
\begin{equation}
\mathcal{H}_{red} = \frac{p_\phi^2}{2a^3} + \frac{a^3}{2} \left( f(Q) -2 Q
f^{\prime }(Q)+2 V(\phi) \right) \approx 0.
\end{equation}

Once more we have the basic Dirac brackets \eqref{exampleDbra} which imply
the consideration of \eqref{canvar} as canonical variables. As far as the
scalar field is concerned, it is not affected by the Dirac brackets since
the only non-zero bracket is $\{\phi,p_\phi\}_D=1$. The reduced total
Hamiltonian in the new canonical variables reads
\begin{equation}  \label{Hredscalar}
\begin{split}
H_{red} & = N \mathcal{H}_{red} +u_N p_N \\
& = N\left(\pm \frac{p_\phi^2}{2 \sqrt{6} \sqrt{Q(q)}p_q }- \sqrt{\frac{3Q(q)%
}{2}} \left(q \mp 2 V(\phi)\right)p_q \right) +u_N p_N,
\end{split}%
\end{equation}
where with $Q(q)$ we denote the inverse of the transformation defined by the
first of \eqref{canvar}. The upper/lower sign in \eqref{Hredscalar}
corresponds to the respective choice for the canonical pair given in %
\eqref{canvar}. The system is still parametrization invariant due to the
arbitrariness of $N$ and the relative multiplier $u_N$. The
reparametrization $N\mapsto n$ with
\begin{equation}  \label{Nrep}
N = \pm 2\sqrt{6} \sqrt{Q} p_q n
\end{equation}
(the upper and lower signs are used respectively in conjunction with those
appearing in \eqref{Hredscalar} and in \eqref{canvar}) leads us to the
reduced Hamiltonian
\begin{equation}
H_{red} = n \bar{\mathcal{H}}_{red} + U_n p_n,
\end{equation}
where $U_n$ is now the new arbitrary multiplier corresponding to $n$. Notice
that in the original variables, the reparametrization \eqref{Nrep}
corresponds to the change $N= 2 a^3 n$. The respective Hamiltonian
constraint in this parametrization is
\begin{equation}  \label{quadcon}
\bar{\mathcal{H}}_{red} = p_\phi^2 \mp 6 (q \mp 2 V(\phi))Q(q) p_q^2 \approx 0 .
\end{equation}

As we observe from \eqref{quadcon}, the parametrization invariance allowed
us to bring the constraint in quadratic form, i.e.
\begin{equation}
\bar{\mathcal{H}}_{red} = G^{IJ}p_I p_J .
\end{equation}
The resulting minisuperspace metric in the coordinates $(\phi,q)$ is
\begin{equation}  \label{minimet}
G_{IJ} =
\begin{pmatrix}
1 & 0 \\
0 & \frac{\mp 1}{6\left(q\mp 2 V(\phi)\right)Q(q)}%
\end{pmatrix}%
.
\end{equation}
For the quantization we may use $\widehat{\bar{\mathcal{H}}}_{red} \Psi =0$ as our
Wheeler-DeWitt equation with the factor ordering for the Hamiltonian
operator being addressed by the Laplacian, i.e.
\begin{equation}
\widehat{\bar{\mathcal{H}}}_{red} = \frac{1}{\mu} \partial_I \left( \mu G^{IJ}
\partial_J \right) ,
\end{equation}
with the measure function being $\mu=\sqrt{|\det G_{IJ}|}$.


In the special case where $V(\phi)$ is constant, i.e. a massless
scalar field in the presence of a cosmological constant, $V(\phi)=\Lambda$,
the $G_{IJ}$ of \eqref{minimet} describes a flat two dimensional space.
Subsequently, the solution to $\widehat{\bar{\mathcal{H}}}_{red} \Psi =0$ can be
expressed in a plane wave-like form as
\begin{equation}  \label{psiflat}
\Psi(\phi,q) = C \exp \left[\mathrm{i} k \phi + k \int\! \sqrt{\frac{\mp 1}{%
6 Q(q) \left(q \mp 2\Lambda\right)}}dq \right],
\end{equation}
where $k$ is the coupling constant and $C$ the normalization constant.
Solution \eqref{psiflat} describes the wave function of any $f(Q)$ theory
with a scalar field, under the assumption $V(\phi)=\Lambda$. The $f(Q)$
theories in this context correspond to different coordinate systems of the flat metric $G_{IJ}$ in which the plane wave solution
\eqref{psiflat} is expressed. For example, in $f(Q)=Q^\lambda$ theory, we
have $q$ related to $Q$ through \eqref{qtoQex1}. For $\lambda >\frac{1}{2}$
we take the upper sign in the \eqref{psiflat}, while for $\lambda <\frac{1}{2%
}$ the lower (these are the appropriate sectors in order for the passing
from one set of variables to the other is done through real
transformations). The integral in the exponent of \eqref{psiflat} then
becomes
\begin{equation}
\int\! \sqrt{\frac{\mp 1}{6 Q(q) \left(q \mp 2\Lambda\right)}}dq= \pm  \sqrt{%
\frac{\mp2(q\mp 2\Lambda)}{3}} \frac{|1-2\lambda|^{\frac{\lambda}{2}}q^{1-\frac{1}{2 \lambda }}}{%
2\Lambda} {}_2F_1(1,\frac{3}{2}-\frac{1}{2\lambda};\frac{3}{2}; 1\mp\frac{q}{%
2\Lambda}),
\end{equation}
where ${}_2F_1$ stands for the Gauss hypergeometric function. The dependence
in the original variable $Q$ is obtained by inverting \eqref{qtoQex1} and
substituting $q = (2\lambda -1) Q^\lambda$ in the expression with the upper
sign or $q = (1-2\lambda) Q^\lambda$ if we take the lower signs.

The case of \eqref{quadcon} is what resembles the most a typical minisuperspace quantization scenario since it leads to a second order partial differential equation. However, there are still important differences, mainly in what regards the configuration space of the problem. In the usual Wheeler-DeWitt quantization, the basic variable is the scale factor $a$, while here this role is reserved for $q$, which is related to the Hubble function.

A fundamental criterion which decides the success of a quantum description is if at some limit we obtain an agreement with the classical solution. In the previous section regarding the vacuum case, where the classical solution corresponds to a constant $q$, we saw that truly the $q=q_{\text{classical}}$ values dominate the probability amplitude $\rho_\Psi(q)=\mu \Psi^*\Psi$. In the presence of matter however, the classical solution  corresponds to a generally dynamical $q(t)$ that changes with time. Hence, we cannot interpret peaks in the probability in the same manner as before. For quantum cosmology, Hartle \cite{Hartlebook} proposed that if the wave function is sufficiently peaked in some region of the configuration space, this implies correlations between observables in this region. It has been argued that these correlations (i.e. relations between positions and momenta) are present when the wave function presents an oscillatory behaviour \cite{Halliwell} in the corresponding variables. It has also been demonstrated that this can serve as a selection rule through symmetries in quantum cosmology, see \cite{Cap}.

As we previously mentioned, our case is pretty distinct, mainly because the corresponding  variables $q$ and $p_q$ are conjugate with respect to Dirac brackets and not with respect to the usual Poisson brackets that are used in non-relativistic quantum mechanics. Thus, we cannot say that a direct analogy can clearly be drawn. However, in the example we are examining here, the situation appears somewhat simpler in deriving some interesting conclusions regarding the correspondence to the classical behaviour.  In what regards the variable, $\phi$, we directly observe from eq. \eqref{psiflat} that the wave function has a purely oscillatory behaviour in the scalar field (as long as we are talking for a real field), which suggests a complete identification with the classical trajectory in $\phi$. For the dependence in $q$, we can make the following observation: First of all let us write the classical solution corresponding to our example $f(Q)=Q^\lambda$ and $V(\phi)=\Lambda$,
\begin{subequations}\label{classolVQ}
\begin{align}\label{classolVQ1}
  N^2 =& 6 (2 \lambda -1)^{\frac{1}{\lambda} } a^{\frac{6}{\lambda }-2}  \left(2 \Lambda  a^6+\kappa\right)^{-\frac{1}{\lambda} }\dot{a}^2  \\ \label{classolVQ2}
  \dot{\phi}^2 = &   6 \kappa (2 \lambda -1)^{\frac{1}{\lambda} } a^{\frac{6}{\lambda }-8} \left(2 \Lambda  a^6+\kappa\right)^{-\frac{1}{\lambda}} \dot{a}^2 .
\end{align}
\end{subequations}
In the above expressions, $\kappa$ is a constant of integration and $a=a(t)$ remains an arbitrary function with respect to which the rest of the degrees of freedom are given. So basically, $a$ serves as a ``time'' variable in the above expressions (for simplicity we may identify $a$ with the variable $t$, i.e. choose the gauge $a(t)=t$). Equation  \eqref{classolVQ2} can be integrated to obtain the $\phi(a)$ in terms of a Gauss hypergeometric function, but we refrain of giving the expression here. By setting the expressions of $f(Q)$, $Q$ and the use of \eqref{classolVQ1} in \eqref{canvar} we see that at the classical level we obtain:
\begin{equation} \label{classqexphi}
  q = \pm \left(2 \Lambda + \frac{\kappa}{a^6}\right).
\end{equation}
Thus, we see that asymptotically, as $a\rightarrow +\infty$, the variable $q$ tends to a constant value $q = \pm 2 \Lambda$. Interestingly enough, this asymptotic ``future'' of the classical solution dominates the probability amplitude $\rho_\Psi(q)$. To see this, let us consider $\lambda>\frac{1}{2}$, $\Lambda>0$ and $\kappa>0$ in \eqref{classolVQ}, so that we do not have to worry about signature changes and the $\phi$ remains real, while $a$ takes values from zero to infinity. For $\lambda>\frac{1}{2}$ the classical behaviour of $q$ is given by the upper sign of \eqref{classqexphi}, so $q>2\Lambda$. The probability amplitude is
\begin{equation}
  \rho_\Psi(q) = \mu \Psi^*\Psi \propto \frac{(2 \lambda -1)^{\frac{1}{2\lambda} } }{ q^{\frac{1}{2\lambda} }(q-2 \Lambda )^{\frac{1}{2}}}
\end{equation}
since both the dependence of the wave function on $\phi$ and $q$ becomes that of a pure phase. The classical domain of definition of $q\in (2\Lambda,+\infty)$, is identified with the quantum one since we need the measure function to be positive. We observe that as $q\rightarrow 2\Lambda^+$ the amplitude diverges. Hence,  at least for the values of the parameters that yield a physically reasonable classical solution, we again obtain a quantum behaviour that agrees at some limit with our observations. It is interesting to note that, in the limit of $q\rightarrow +\infty$, which corresponds to the classical singular value $a\rightarrow 0$, we obtain $\rho_\Psi \rightarrow 0$, which can be interpreted as some form of resolution for the classical singularity.

Of course the situation for the general problem is quite more complicated and we cannot comment that a similar correspondence will always be achieved in all physically relevant configurations.

\section{Perfect fluid source}

\label{sec7}

In this section we see how the theory changes when a perfect fluid is
considered as a matter source. In order to write a minisuperspace Lagrangian
we need to assume a particular form for the equation of state connecting the
pressure $P$ with the energy density $\rho$. The simplest assumption is of
the form $P=w \rho$, with $w$ being constant. The energy momentum tensor in
mixed components is of course $T^{\mu}_{\phantom{\mu}\nu}=\mathrm{diag}%
(-\rho,P,P,P)$. If we take the time derivative of the constraint equation of %
\eqref{feqgrav} (the pure temporal (00) component) and use also the rest of %
\eqref{feqgrav} it is easy to derive the typical continuity equation \cite%
{Xu}
\begin{equation}
\dot{\rho} + 3 \frac{\dot{a}}{a} \left( \rho+P\right) =0 .
\end{equation}
This is related to the fact that the field equation for the connection
vanishes identically for an FLRW spacetime \cite{Koivisto1}. The linear form
of the equation of state then implies: $\rho = -\rho_0 a^{-3(w+1)}$, where $%
\rho_0$ is a constant of integration. This results in the following matter
minisuperspace Lagrangian
\begin{equation}
L_m = \sqrt{-g} \rho = - \rho_0 N a^{-3w} .
\end{equation}
By using the latter into \eqref{Lag} it is easy to verify that the emerging
Euler-Lagrange equations are equivalent to the field equations of $f(Q)$
gravity \eqref{feqgrav}, where in the right hand side the matter content is
given by an energy momentum tensor we mentioned earlier.

Exactly as before, the total Hamiltonian is given by \eqref{totHamsc}, where
now the Hamiltonian constraint is
\begin{equation}
\mathcal{H} = -\frac{p_a^2}{12a f^{\prime }(Q)} + \frac{a^3}{2} \left(
f(Q)-Q f^{\prime }(Q)\right) + \rho_0 a^{-3 w} .
\end{equation}
We have the same basic quadruplet of constraints as in the previous cases $%
p_N\approx 0$, $p_Q \approx 0$, $\mathcal{H}\approx 0$ and $\chi \approx 0$,
where $\chi$ is again given by \eqref{chicon}. As in the scalar field case,
we need to consider a different combination of constraints in order to
reveal the maximum number of first class quantities, this is: $\zeta_1 = p_N$%
, $\zeta_2 = p_Q$, $\zeta_3 = \chi$, $\zeta_4 = \mathcal{H} - \frac{p_a
\left(f(Q)-2 Q f^{\prime }(Q)-\rho_0 (w-1) a^{-3 (w+1)}\right)}{a^2
f^{\prime }(Q) \left(2 Q f^{\prime \prime }(Q)+f^{\prime }(Q)\right)} p_Q$.
The $\zeta_1$ and $\zeta_4$ commute weakly with all the constraints, while
the commutator between $\zeta_2$, $\zeta_3$ is given by \eqref{secondclass}.
The conservation of $\chi$ this time fixes the velocity to be
\begin{equation}
u_Q \approx \frac{N p_a a^{-3 w-5} \left(2 \rho_0 w-a^{3( w+1)} \left(f(Q
)-2 Q f^{\prime }(Q)\right)\right)}{2 f^{\prime }(Q) \left(2 Q f^{\prime
\prime }(Q )+f^{\prime }(Q )\right)} .
\end{equation}

After the introduction of the Dirac brackets and the use of the second class
constraints as strong equations $\chi=0$ and $p_Q=0$ the reduced Hamiltonian
constraint is written
\begin{equation}
\mathcal{H}_{red} = \frac{a^3}{2} \left(f(Q)-2 Q f^{\prime }(Q)\right) +
\rho_0 a^{-3 w} \approx 0.
\end{equation}
The same line of reasoning as before is followed. We have the basic
canonical variables, as imposed by the Dirac bracket \eqref{Dbracan}, that
we see in \eqref{canvar}. We introduce once more the function $F(q)$ from %
\eqref{newparq} and express the Hamiltonian constraint in the new variables
as
\begin{equation} \label{priorparH}
\mathcal{H}_{red} = -\frac{p_q}{F(q)^2} + \frac{\rho_0}{2^w} \frac{q^w
F(q)^{2w}}{|p_q|^w} \approx 0 .
\end{equation}
The absolute value over $p_q$ serves to cover both possibilities regarding
the canonical pair \eqref{canvar}. Given once more the parametrization
invariance of the system through the mapping $N\mapsto n=\frac{q^{w}F^{2w}}{%
|p_{q}|^{w}}N$ in the reduced Hamiltonian $H_{red} = N \mathcal{H}_{red} + u_N p_N$, we obtain
\begin{equation}
H_T = n \bar{\mathcal{H}}_{red} + U_n p_n
\end{equation}
with
\begin{equation}
\bar{\mathcal{H}}_{red}=\mp \frac{|p_{q}|^{1+w}}{q^{w}F(q)^{2(1+w)}}+\frac{\rho
_{0}}{2^{w}}\approx 0.  \label{conredflu2}
\end{equation}
As we see from the above constraint, the $w=-1$ case corresponds to $q(t) =
\pm 2 \rho_0$. Thus, we obtain $Q(t)=$constant, irrespectively of the $f(Q)$
theory that one may consider.

The quantum realization of the classical constraint \eqref{conredflu2} is not
at all trivial. This is due to the momentum $p_q$ being raised to a power
that is not necessarily an integer number. We distinguish the
simpler cases of dust ($w=0$) and rigid matter ($w=1$). In the generic case, where $1+w$ is non-integer this type
of situation is being studied in the context of fractional Quantum
Mechanics \cite{Laskin1,Laskin2,Laskinbook}. The idea behind the theory is
the replacement of the Brownian paths in the Feynman path integral formulation of nonrelativistic Quantum Mechanics
by L\'evy flights \cite{Laskinbook}. The fractal dimension $\alpha$ of the
L\'evy flight appears as the power of the momenta, $|p|^\alpha$, in the
Hamiltonian with $1<\alpha\leq 2$. In our case we have $\alpha=1+w$. The $%
\alpha=2$ case corresponds to Brownian paths and the usual quantum
mechanics. For the quantum version of $|p|^\alpha$ the Riesz fractional
derivative is used, which is defined for one dimensional problems as
\begin{equation}  \label{RieszD}
(\hbar \nabla)^{\alpha} \psi(q) = -\frac{1}{2\pi \hbar} \int_{-\infty}^{+%
\infty} \!\!  e^{\frac{\mathrm{i} p q}{\hbar}}|p|^\alpha \phi(p) dp,
\end{equation}
where $\phi(p)$ is the Fourier transform of $\psi(q)$ given by
\begin{equation}
\phi(p) = \int_{-\infty}^{+\infty} e^{-\frac{\mathrm{i} p q}{\hbar}}
\psi(q) dq.
\end{equation}

Nevertheless, there exist other possibilities for the quantum analogue of %
\eqref{conredflu2}, involving different definitions of a fractional derivative. In \cite{Katugampola} a more ``easily manageable''
fractional derivative was introduced, whose possible uses in quantum
mechanics have been stressed in \cite{DAnderson}. It is defined as
\begin{equation}
  D^\alpha (\Psi(q)) = \lim_{\epsilon\rightarrow 0}\frac{\Psi(q e^{\epsilon q^{-\alpha}}) - \Psi(q)}{\epsilon}, \quad 0<\alpha \leq 1, \quad q>0 .
\end{equation}
Based on the above definition, the action of this derivative on differentiable functions yields
\begin{equation}  \label{fracdif1}
D^\alpha (\Psi(q)) = q^{1-\alpha} \frac{d\Psi}{dq}
\end{equation}
when $0<\alpha\leq 1$. On the other hand, if $1<\alpha\leq 2$ the expression
\begin{equation}  \label{fracdif2}
D^\alpha (\Psi(q)) = q^{2-\alpha}  \frac{d^2\Psi}{dq^2}
\end{equation}
is used instead. In general, for $\alpha \in (n,n+1]$ for some $n\in \mathbb{N}$, we have the generalization $D^\alpha = q^{n+1-\alpha}  \frac{d^{n+1}}{dq^{n+1}}$ \cite{Katugampola}.

Of course, the derivative \eqref{fracdif1} cannot be simply used in place of $|p|^\alpha$, since we need to also demand the Hermiticity of the relevant operator. By using the general expression for a Hermitian operator, as also presented in \cite{DAnderson}, we write for \eqref{conredflu2} the following Hamiltonian operator assuming $0<\alpha\leq 1$,
\begin{equation}\label{Hamfrac}
   \widehat{\bar{\mathcal{H}}}_{red}\Psi = \pm \frac{\ima}{2\mu(q)} \left[ \mu(q) A(q) q^{1-\alpha} \partial_q \Psi  \; +\partial_q \left(\mu(q) A(q) q^{1-\alpha}\Psi \right) \right] + k^2 \Psi
\end{equation}
where $A(q) = q^{-w} F(q)^{-2(1+w)}$, $\alpha=1+w$, $k^2=2^{-w}\rho_{0}$ and $\mu(q)$ the measure function. The quantum Dirac proposition of the quantum constraint annihilating the wave function leads to the equation $\widehat{\bar{\mathcal{H}}}_{red} \Psi =0$, which results in
\begin{equation}
  \begin{split}
    - \frac{q^{-2 w-1}}{2 \mu (q) F(q)^{2 (w+1)} } \left[q \Psi (q) \mu '(q)+\mu (q) \left(2 q \Psi '(q)-2 w \Psi (q)\right)\right]+ \frac{(w+1)\Psi (q)  F'(q)}{ q^{2 w}F(q)^{2 w+3}} = \pm \ima k^2 \Psi(q),
  \end{split}
\end{equation}
with the solution
\begin{equation} \label{Hamfracsol}
  \Psi(q) = C \frac{q^w F(q)^{1+w}}{\sqrt{\mu(q)}} \exp \left(\pm \ima k^2 \int\!\! q^{2w} F(q)^{2(w+1)} dq \right),
\end{equation}
where once more $C$ denotes the normalization constant. Notice that this result is compatible with the solution \eqref{psigen} of the vacuum case when $w=k=0$ (remember that $k=0$ implies $\rho_0=0$).

As we stated, equation \eqref{Hamfrac} refers to the case $0<\alpha\leq 1$, which corresponds to $-1<w\leq 0$. In order to utilize \eqref{fracdif2}, which is relevant for $1<\alpha\leq 2$ (or equivalently $0<w\leq 1$) in the context of a Hermitian operator, we write an expression that resembles a one-dimensional Laplacian, so that for the Hamiltonian operator we have
\begin{equation} \label{Hamfracquad}
  \widehat{\bar{\mathcal{H}}}_{red}\Psi = \pm \frac{1}{\mu(q)} \partial_q \left[ \mu(q) A(q) q^{2-\alpha} \partial_q \Psi \right] + k^2 \Psi =0 .
\end{equation}
where in accordance with the usual Laplacian, the measure function is $\mu(q) = A(q)^{-\frac{1}{2}}q^{1-\frac{\alpha}{2}}$. By taking into account the previous relations  for $A(q)$ and $\alpha$ the general solution to the above equation is written in the form:
\begin{equation} \label{solHamquad}
  \Psi = C_1 \exp \left[ \sqrt{\mp k^2} \left(\int\!\! q^{w-\frac{1}{2}} F(q)^{w+1} dq \right) \right] + C_2 \exp \left[ - \sqrt{\mp k^2} \left(\int\!\! q^{w-\frac{1}{2}} F(q)^{w+1} dq \right) \right],
\end{equation}
where $C_1$ and $C_2$ are the integration constants. For the extreme value $\alpha=2$, corresponding to $w=1$, we acquire the solution to the Laplacian as we should expect from the definition we used for the operator \eqref{Hamfracquad}. By looking at \eqref{solHamquad}, we observe that, for a positive energy density $\rho_0>0$, which implies $k^2>0$, and for the upper sign of \eqref{Hamfracquad} the solution represents combinations of ingoing and outgoing waves, e.g. $\Psi_k= C \exp\left(\pm \ima k\int\!\! q^{w-\frac{1}{2}} F(q)^{w+1} dq\right)$ - given of course that we have assigned variables so that the rest of the involved quantities, $q$, $F(q)$, are positive. For an ingoing or an outgoing such wave, if we also adopt a change of variables $x=\int\!\! q^{w-\frac{1}{2}} F(q)^{w+1} dq$ we can write
\begin{equation}
  \int \mu(q) \Psi_{k'}^*(q) \Psi_k(q) dq = \int \!\! q^{w-\frac{1}{2}} F(q)^{w+1}  \Psi_{k'}^*(q) \Psi_k(q) dq =|C|^2 \int e^{\ima (k-k') x} dx,
\end{equation}
which for a domain of definition $x \in \mathbb{R}$ is normalizable to a delta function.

By comparing solutions \eqref{Hamfracsol} and \eqref{solHamquad} we observe that this distinction of operators based on $\alpha$ creates a discontinuity at $\alpha=1$ or equivalently at $w=0$. The limit \eqref{solHamquad} as $w\rightarrow 0^+$ does not recover expression \eqref{Hamfracsol}.

Lastly, we want to comment on what can be done if we need to consider $w<-1$. In that case, we can go back in \eqref{priorparH} and instead of adopting the parametrization that lead us to \eqref{conredflu2}, assume a change of the form $N\mapsto \tilde{n} = \frac{2^w|p_q|}{\rho_0 F(q)^2} N$ which gives rise to the constraint
\begin{equation}
\tilde{\mathcal{H}}_{red}=q^{w}F(q)^{2(1+w)} |p_{q}|^{-(1+w)} \mp \frac{2^{w}}{\rho
_{0}}\approx 0.  \label{conredflu3}
\end{equation}
The latter can be treated in a similar manner as we have done with \eqref{conredflu2}.

\section{Conclusions}

\label{con}

We used Dirac's method of quantizing constrained systems in the context of $f(Q)$ cosmology in the coincident gauge. We demonstrated that the resulting theory contains second class constraints. Thus, the quantization of $f(Q)$ cosmology goes outside the usual straightforward quantization that we are accustomed to from theories where only first class constraints are present.

The second class constraints correspond to redundant degrees of freedom. In order to eliminate them we introduced the Dirac brackets and found appropriate variables for constructing a canonical quantization scheme. As a result, we obtained that the relative quantum theory of $f(Q)$ cosmology is highly dependent on the nature of the matter content: in the vacuum case a linear in the ``momentum'' Hamiltonian emerged; for a scalar field, with the help of the parametrization invariance, we were able to write a typical quadratic expression; finally for a perfect fluid, we obtained something that even goes outside the scope of the usual quantum mechanics requiring a generalization of the theory in terms of fractional derivatives.

We managed to derive the generic expressions for the wave function for every case under study. Thus, obtaining a result for every possible $f(Q)$ modification of the classical theory. To our knowledge this is the only work up to now that takes into account the singular nature of the system and the existence of second class constraints in the process of quantization. Based on our results, we saw that, in the vacuum case, the classical solution strongly dominates at the quantum level; even though it is not the single possibility and other configurations are also permissible. This property also became evident through some simple examples we considered. The situation when matter comes into play becomes a lot more complicated and further studies regarding particular models need to be performed. In the case of the scalar field the choice of a potential $V(\phi)$ and $f(Q)$ governs the geometry of the reduced minisuperspace which we obtained, while in the case of a simple perfect fluid there exist a lot of ambiguities even at the level of the operators that one might use. What is more interesting, is the emergence of the need for the theory of fractional Quantum Mechanics in the description of the system. In a recent work \cite{Jalal}, a connection between quantum modifications in terms of fractional derivatives in the Hamiltonian constraint operator and generalizations of the black hole entropy, like the Barrow entropy \cite{BarrowE} (for further applications see \cite{BarrowE2,BarrowE3,BarrowE4,BarrowE5}), has been explored. It is rather intriguing that in the case of a perfect fluid in non-linear $f(Q)$ theory, such a modification appears in a natural manner already from the classical constraint. For further applications of fractional quantum mechanics in a cosmological context see \cite{Moniz1,Moniz2}.

We hope that this work will motivate further studies on this subject, since it seems that the $f(Q)$ theory allows for significant deviations from the usual applications of quantum cosmology in General Relativity,  $f(R)$ gravity or other theories where only first class constraints are present.

Finally, it is important to mention here that the results of this study are also applicable in the case of $f(T)$ teleparallel cosmology, since $f(T)$ and $f(Q)$ theories share the same dynamics in the cosmological background space.

\section*{Acknowledgements}
N. D. acknowledges the support of the Fundamental Research Funds for the Central Universities, Sichuan University Full-time Postdoctoral Research and Development Fund No. 2021SCU12117.

\end{document}